# High ferromagnetic phase transition temperatures in GaMnAs layers annealed under arsenic capping

J. Sadowski[1,2], J. Z. Domagała[1], V. Osinniy[1], J. Kanski[3], M. Adell[3], L. Ilver[3], C. Hernandez[4], F. Terki[4], S. Charar[4] and D. Maude[5].

[1] Institute of Physics, Polish Academy of Sciences, al Lotników 32/46, 02-668 Warszawa, Poland

[2] Institut für Angewandte und Experimentelle Physik, Universität Regensburg, 93040 Regensburg, Germany

[3] Department of Experimental Physics, Chalmers University of Technology, SE-41296 Göteborg, Sweden

[4] Groupe d'Etude des Semiconducteurs CC074, Université Montpellier II, Montpellier, France

[5] High Magnetic Field Laboratory, CNRS-MPI, 25 Avenue des Martyrs, 38042 Grenoble, France

**Abstract**

Thin GaMnAs layers were grown by molecular beam epitaxy and were subjected to low-temperature post growth annealing under an arsenic capping layer. Modifications of the magnetic, transport and structural properties due to the post-growth annealing are presented. It is shown that the presence of arsenic capping significantly shortens the annealing time for complete removal of Mn interstitials from the GaMnAs volume. As a result the concentration of free carriers is increased, and the ferromagnetic phase transition temperature is significantly enhanced. The GaMnAs layers annealed in this way can be further overgrown by epitaxial films.

## 1. Introduction

GaMnAs is a model semiconductor with carrier induced ferromagnetism. It is currently being used to test the prototype spintronic devices, such as spin-diodes [1], giant magnetoresistance structures [2]. Since the first report on ferromagnetism in GaMnAs, by H. Ohno et al. [3], a lot of interesting, spin-related phenomena have been observed in this material. The remarkable magnetotransport properties of GaMnAs enabled observation of a giant planar Hall effect [4], current driven magnetization reversal [5], dependence of a magnetic anisotropy on the concentration of carriers [6]. In spite of these results associated with the magnetic properties of GaMnAs, it is still desirable to improve its quality and obtain a material with higher ferromagnetic phase transition temperature (Tc), limited nowadays to 173 K. The mean field, Zener model of carrier mediated ferromagnetism in transition metal doped semiconductors developed by T. Dietl et al. [7] predicts Tc to reach the room-temperature for GaMnAs containing 10% Mn [8], and sufficiently high concentrations of carriers (valence band holes). Recently the ferromagnetism persisting up to 250 K has been reported [9] in AlGaAs/GaAs two-dimesional hole gas structures with Mn delta doping. The Tc increase in the uniform GaMnAs films was achieved as a consequence of developing efficient methods leading to removal of the prevailing compensating defects - Mn in the interstitial positions ($Mn_I$), which, as shown both theoretically [10, 11] and experimentally are always present in the as-grown GaMnAs, with concentrations of up to about 20% of the total Mn content. As it was first demonstrated by Edmonds et al. [12], the $Mn_I$ defects can be removed by a low temperature post-growth annealing of the samples exposed to the air after the MBE growth. The annealing temperatures used in the annealing processes can be lower than the GaMnAs growth temperature, and then extremely long annealing times (up to 100 hrs) are used [12], however higher annealing temperatures and short annealing times (0.5 – 3 hrs) were successfully applied by other groups [13 -15]. The low annealing temperatures are the implication of a metastable character of GaMnAs. At temperatures above 300 °C it starts to decompose to MnAs inclusions inside the GaAs matrix. These inclusions, if sufficiently large are also ferromagnetic, since MnAs is a ferromagnetic metal with Tc of 40 $^{o}$C. However, the GaAs:MnAs system is usually highly resistive, and the main advantage of GaMnAs – namely ferromagnetism mediated by carriers in a semiconductor, is lost in such a case. In consequence the annealing temperatures and annealing times should be carefully chosen to remove Mn from interstitial positions, but not to decompose the GaMnAs ternary alloy. In this paper we present a modification of the low temperature annealing method, which uses amorphous As layers deposited on GaMnAs surface directly after the MBE growth as a passivating medium for out-diffused Mn interstitials. So far such passivation has been achieved by oxidation. We show that our method is more efficient in terms of shorter annealing times and lower annealing temperatures than annealing in air. Moreover it leaves GaMnAs surface suitable for further epitaxy, either in the same MBE system used for GaMnAs layer growth, or in another MBE system, since As capping also forms a protective layer inhibiting the surface oxidation and preserving clean, atomically flat, as-grown surface.

## 2. Sample preparation

The GaMnAs samples were grown in the KRYOVAK MBE system, located at the Swedish synchrotron radiation centre MAX-Lab in Lund, Sweden. Arsenic was supplied from a valved cracker source decomposing $As_4$ to $As_2$ molecules. The layers were grown on epiready GaAs(100) wafers, which were In-glued to molybdenum holders. After thermal desorption of native oxide from the GaAs substrate, a standard, high temperature GaAs buffer was deposited. The substrate temperature was then decreased to 230 – 240 $^{o}$C, and a GaMnAs layer was grown with low $As_2$ flux ($As_2$ to Ga flux ratio was about 2) and the growth rate of about 0.2 molecular layer (ML) per second. The growth was monitored by reflection high energy electron diffraction (RHEED) system. A diffraction pattern of a two-dimensional monocrystalline GaMnAs surface was observed, with (1x2) surface reconstruction, and without any admixture of any other phase (MnAs inclusions). Oscillations of the specular beam intensity were used to measure the GaMnAs composition in-situ, by measuring the growth rate increase of GaMnAs, with respect to GaAs. The thicknesses of GaMnAs layers were chosen between 100 and 1000 Å, since thicker layers usually exhibit worse magnetic properties. After completing the GaMnAs growth the As flux was stopped by closing the As cracker valve and the shutter. Then the substrate heater was turned off, and when the temperature fell below 150 $^{o}$C, the As flux was opened, and a thick amorphous As layer was deposited. The thickness of this capping layer, was in the range of 1000 – 2000 Å as estimated from SIMS.

## 3. Defects in GaMnAs

The most abundant defects in GaMnAs are As antisites ($As_{Ga}$) an Mn interstitials. Both defects are double donors, effectively compensating some part of Mn acceptors, (i.e. Mn atoms located at Ga sites ($Mn_{Ga}$), which are single acceptors in the GaAs host lattice. The Tc in GaMnAs is proportional to the concentration of $Mn_{Ga}$ and to the concentration of valence band holes. In the Zener mean field model [7], the dependence of Tc is described by the relation::

$Tc \sim [Mn_{Ga}] \cdot p^{1/3}$     (1),

where p is the concentration of valence band holes. Taking the compensating defects into account, p is given by

$p = [Mn_{Ga}] - 2([As_{Ga}] + [Mn_I])$    (2)

According to (1) and (2) maximizing Tc requires both maximum concentration of $Mn_{Ga}$ and minimum density of $As_{Ga}$ and $Mn_I$ defects. Unfortunately these two demands are in practice contradictory, since increasing the Mn content in GaMnAs requires lowering the growth temperature [16, 17], and the low growth temperature promotes formation of As antisite defects [18]. It means that there might be some optimum Mn composition, at which these two tendencies compromise. This optimum Mn composition was found so far to be in the range of 6% - 8%. The maximum Mn content achieved until now in uniform GaMnAs is 10% [19], although growing $Ga_{0.9}Mn_{0.1}As$ requires very low growth temperatures (about 160 $^o$C) and special growth procedures. The concentration of the compensating defects ($As_{Ga}$ and $Mn_I$) can reach levels as $[As_{Ga}]_{max} \approx 0.5$ at%, $[Mn_I]_{max} \approx 0.17$ $[Mn_{Ga}]$ [20, 21]. As a consequence the compensation degree in GaMnAs can be as high as 50% - 70%. Indeed such high compensation levels are observed in as-grown GaMnAs. As an example we show in Fig.1 the results of the Hall effect measurements for a 300 Å thick $Ga_{0.94}Mn_{0.06}As$. The usual problem in extracting carrier concentrations from the Hall effect measurements in GaMnAs is a negative magnetoresistance, which can be quite large (20 – 30%). In the present case the negative magnetoresistance is rather small, only about 2%. Therefore the value of p can be obtained from the $R_{Hall}$ vs. B slope without the necessity to subtract the negative magnetoresistance. We thus obtain p= 6.5 x $10^{20}$ cm$^{-3}$. The Mn content of 6% corresponds to a concentration of Mn acceptors of 1.32 x $10^{21}$ cm$^{-3}$, which means that the compensation of $Mn_{Ga}$ in this sample is at the level of 50%. This compensation can be due to both $As_{Ga}$ and $Mn_I$, and it is not possible to estimate their individual contributions without additional measurements. Direct measurements of $Mn_I$ concentrations are rather difficult. The first experimental observation of $Mn_I$ was obtained from Rutheford back-scattering, and particle induced X-ray emission (PIXE) measurements [21]. Substitutional defects like $As_{Ga}$ do not show up in these methods, but as their presence affects the lattice constant, their density can be estimated by X-ray diffraction. The situation is somewhat complicated for GaMnAs, since in this case both $As_{Ga}$ and $Mn_I$ cause the lattice expansion. Nevertheless, it is possible to separate the contributions from these two effects using XRD [22]. Fig.2 shows the results of such studies of three 500 Å thick $Ga_{0.96}Mn_{0.04}As$ samples grown at three different $As_2$:Ga flux ratios 2, 5 and 9. Under these conditions different $As_{Ga}$ concentrations are expected. The inset shows a comparison between measured and calculated XRD spectra for the sample with the lowest As$_2$/Ga flux ratio. The GaMnAs peaks are broadened due to the small film thickness. The presence of X-ray interference fringes in the measured curve, and the very good correspondence between measured and calculated spectrum proves the high quality of the GaMnAs layers. The As$_{Ga}$ concentrations can be evaluated by the angular positions of the Bragg diffraction peak resulting from diffraction in a thin LTGaAs buffer deposited prior to the growth of GaMnAs. Since GaMnAs layers are grown on these buffers at the same growth conditions it can be expected that As$_{Ga}$ concentrations are the same in GaMnAs as in LT GaAs. It is clear that the sample grown at the lowest As$_2$:Ga flux ratio also has the lowest As$_{Ga}$ concentration - the diffraction peak from the LT GaAs buffer appears only as a shoulder at the left side of the main (004) peak from the GaAs substrate. For samples grown at higher As$_2$ to Ga flux ratios, peaks from LT GaAs are shifting more towards lower diffraction angles, which reflects a lattice expansion caused by the As$_{Ga}$ defects. On the other hand, the diffraction peaks resulting from GaMnAs layers are shifting to higher angles with increased As$_2$:Ga flux ratio. This suggests that with increasing concentration of As$_{Ga}$ in GaMnAs, the concentration of Mn$_I$ decreases. A quantitative analysis of this process has been presented elsewhere [22].

**4. Defects removal by the post-growth annealing.**

From the two defects considered above, the As$_{Ga}$ is stable at the relatively low post-growth annealing temperatures, so only the Mn$_I$ can be affected by the treatment [12 – 15, 23]. It is known that the effects of annealing are efficiently suppressed by very thin LT GaAs capping layers [24, 25], and that the annealing efficiency can be increased by applying the treatment in an oxygen atmosphere [26], or by surface roughening [27]. All these result suggest that the post-growth annealing mechanism is associated with diffusion of the Mn$_I$ atoms and trapping them at the surface by chemical bonds (with oxygen or nitrogen in cases when GaMnAs surface is exposed to the air). We proposed recently to use an amorphous As capping as a reactive agent [23]. An important advantage with this procedure is that it can be applied in the MBE growth chamber, immediately after the growth.

Fig. 3 presents results of Hall effect measurements for the same sample as in Fig.1, but after the post growth annealing process performed at 180 °C during 3 hours. The efficiency of removal of Mn interstitials is close to 100%, since the concentration of holes increased from $6.5 \times 10^{21}$ cm$^{-3}$ to the value of $1.3 \times 10^{21}$ cm$^{-3}$. This corresponds almost exactly to the content of $Mn_{Ga}$ which is 6%, as calibrated by RHEED intensity oscillations. The ferromagnetic phase transition temperature for this sample is 150 K.

Fig. 4 shows the temperature dependence of magnetization, measured by a SQUID magnetometer for the sample similar to the one presented in Fig. 3, namely a 400 Å thick $Ga_{0.94}Mn_{0.06}As$ layer with an amorphous As capping, before and after annealing at 180 °C for 2h. A remarkable increase of Tc from about 70 K, to 145 K can be seen also for this sample. The Tc increase is correlated with an increase of a saturation magnetisation, observed at low temperatures. This is due to removal of antiferromagnetically coupled $Mn_{Ga}$-$Mn_I$ pairs, which are excluded from participating in FM phase in the non-annealed sample. In Fig.5, which shows the temperature dependence of magnetisation for 500 Å thick $Ga_{0.94}Mn_{0.06}As$ layer after 6h annealing under As capping at 180 °C, we see that a Tc close to 160 K is reached. The efficiency of removing $Mn_I$ defects by annealing As capped samples is also confirmed by XRD measurements. As explained above, the Mn interstitials cause the GaMnAs lattice to expand, so the lattice constant is expected to decrease with post-growth annealing. This is really the case, as shown in Fig. 6. The figure shows significant shifts of (004) Bragg reflections from a 1000 Å thick $Ga_{0.94}Mn_{0.06}As$ layer after consecutive annealings during 1, 3 and 30 hrs at 180 °C. Already after 1 h annealing the GaMnAs lattice parameter is significantly reduced, annealing for 3 hrs induce only a small further reduction, whereas annealing for 30 hrs has almost no further effect.

## 5. Conclucions

We have presented an efficient post-growth annealing method employing an adsorbed As layer as the agent for trapping out-diffusing Mn interstitials from MBE grown GaMnAs. The efficiency of the method has been proven by measurements of a high density of free carriers (holes) to the range of $10^{21}$ cm$^{-3}$, and a $T_c$ value of up to 160 K for a GaMnAs layer with 6 at% Mn. The present treatment is carried out in the MBE growth chamber, and has the important advantage over other method that the surface of the annealed layer can be restored for further epitaxial growth on top of the high quality, Mn interstitial-defect-free GaMnAs layers.

**Acknowledgements:**

The present work is part of a project supported by the Swedish Research Council VR. One of the authors (J.S) acknowledges the financial support from the Polish State Committee for Scientific Research (KBN) through Grant No PBZ-KBN-044/P03/2001. Measurements at high magnetic fields have been supported by the European Community within the 'Access to Research Infrastructure action of the Improving Human Potential Programme.

**Figure captions**

**Fig.1.** Results of Hall effect measurements for as-grown 300 Å thick $Ga_{0.94}Mn_{0.06}As$ layer. Solid, blue curve – hall resistance $R_{Hall}$, dashed red curve – longitudinal resistance $R_{xx}$. The concentration of holes in the sample is equal to $6.5 \times 10^{20}$ cm$^{-3}$, which means 50% compensation of $Mn_{Ga}$ acceptors.

**Fig.2.** (004) X-ray Bragg reflections for three 500 Å thick $Ga_{0.96}Mn_{0.04}As$ samples grown at different $As_2$/Ga flux ratios. The highest intensity peak on the right side results from diffraction on the GaAs(001) substrate. Broad peaks at lower angles are due to the diffraction on GaMnAs layers. The inset shows the comparison between simulated and measured spectrum for the sample grown at the lowest $As_2$/Ga flux ratio.

**Fig.3.** Results of Hall effect measurements for 300 Å thick $Ga_{0.94}Mn_{0.06}As$ layer annealed for 3h at 180 °C. Solid, blue curve – hall resistance $R_{Hall}$; dashed, red curve – longitudinal resistance $R_{xx}$. The concentration of holes in the sample is equal to $1.3 \times 10^{21}$ cm$^{-3}$, which corresponds almost exactly to the content of Mn at Ga sites, which means that Mn interstitial defects have been effectively removed by the post-growth annealing procedure. The scale for $R_{Hall}$ and $R_{xx}$ is the same as in Fig.1.

**Fig.4.** Temperature dependence of magnetization for 400 Å thick $Ga_{0.94}Mn_{0.06}As$ layer with amorphous As capping before and after the post-growth annealing, at 180 °C for 3h. Post-growth annealing increases both Tc and saturation magnetization.

**Fig.5.** Temperature dependence of magnetization for 500 Å thick $Ga_{0.94}Mn_{0.06}As$ layer with amorphous As capping, after the post-growth annealing, at 180 °C for 6h. Tc value of 160K is the highest reached so far for $Ga_{0.94}Mn_{0.06}As$.

**Fig.6.** (004) X-ray Bragg reflections for non annealed and annealed pieces of As capped 1000 Å thick $Ga_{0.94}Mn_{0.06}As$ layers. The shifts of GaMnAs diffraction peaks towards higher angles after annealing are due to the annealing induced reduction of GaMnAs lattice parameter.

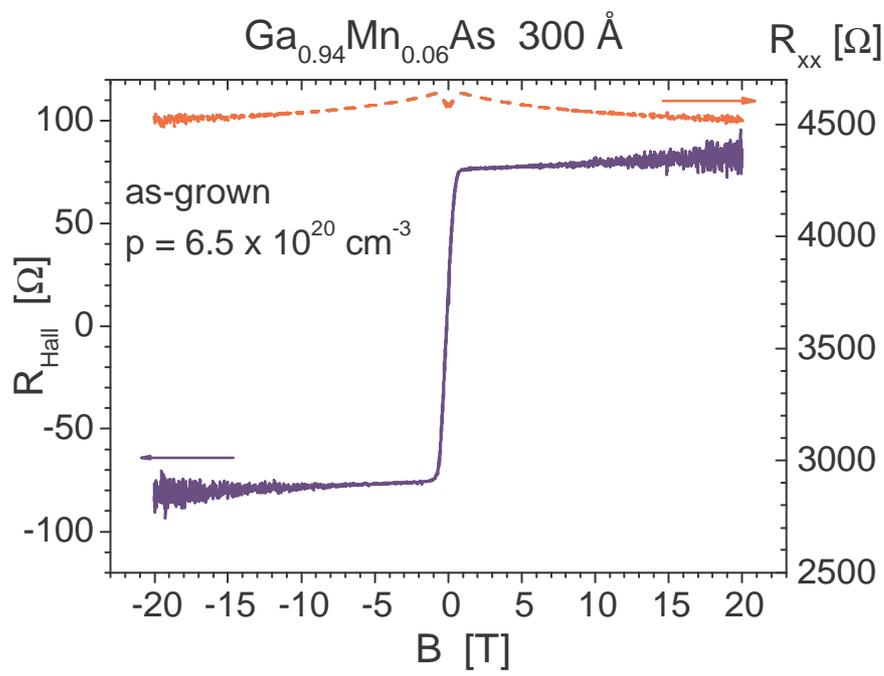

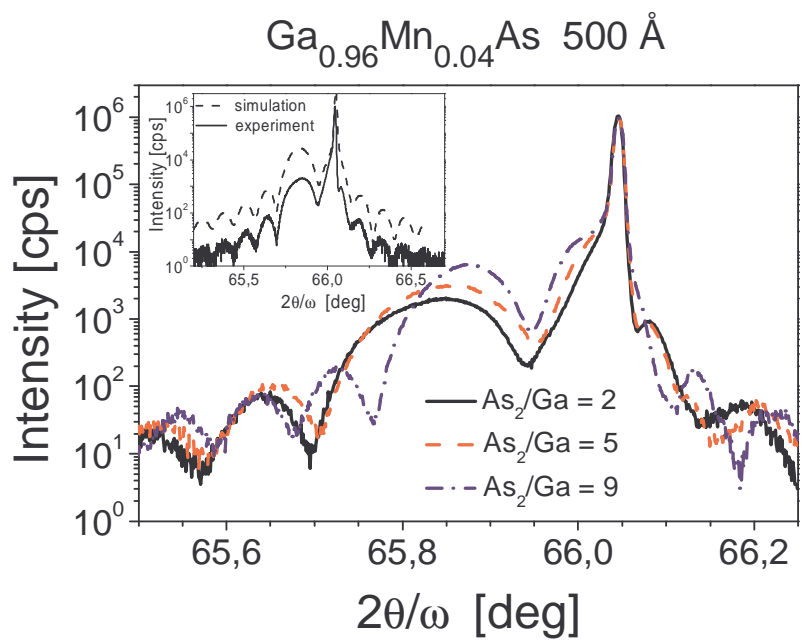

J. Sadowski et al. Fig.2

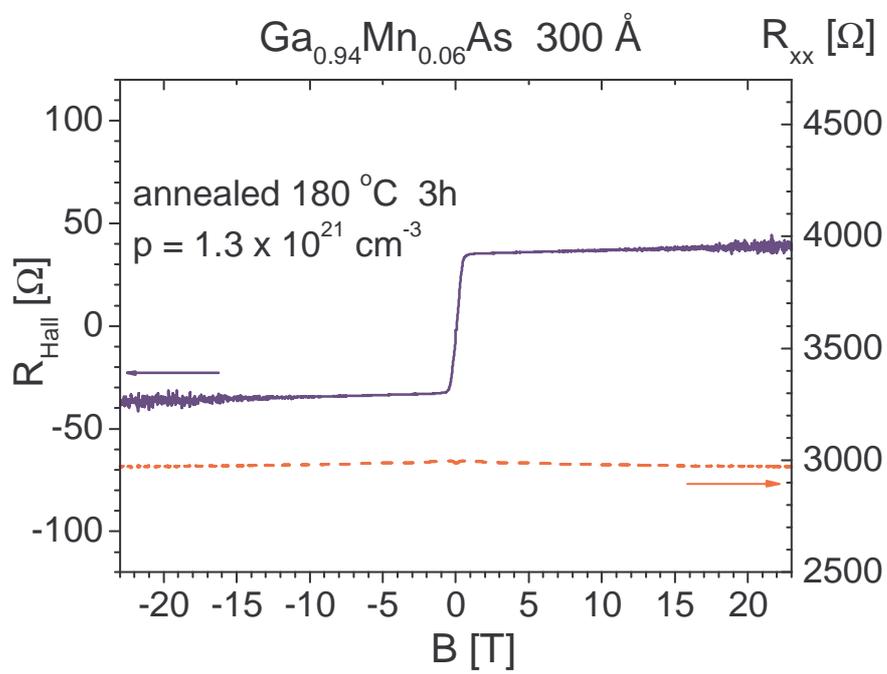

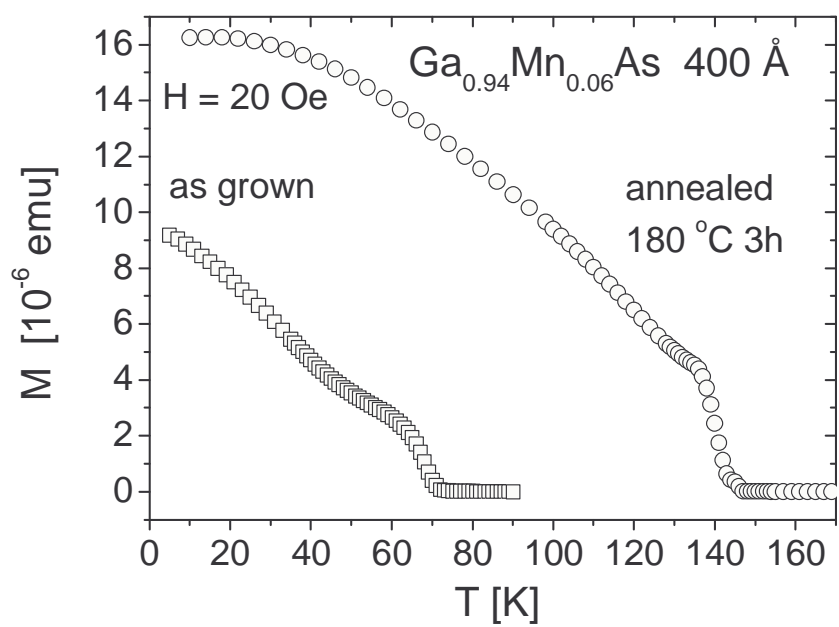

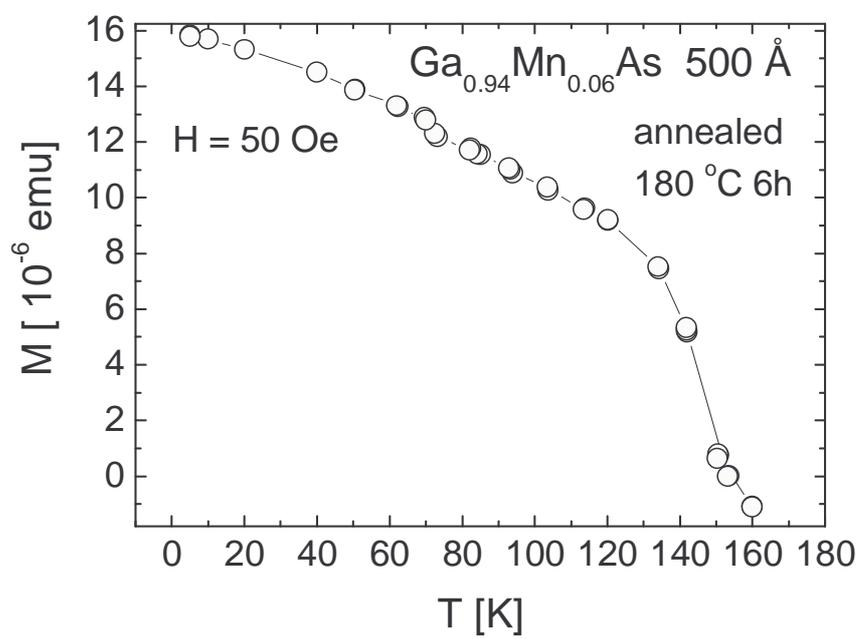

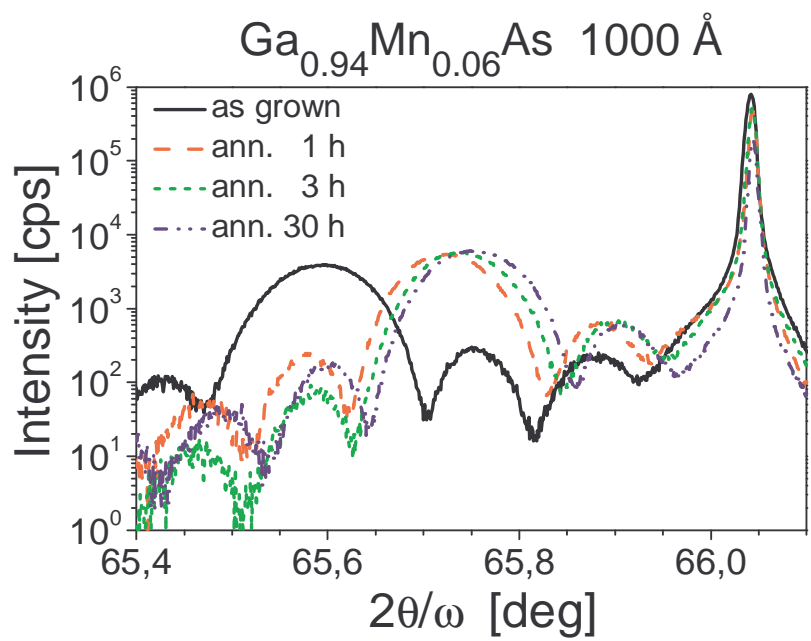

J. Sadowski et al. Fig.6